# Negative Interatomic Spring Constant Manifested by Topological Phonon Flat Band

Bowen Xia,[1] Hang Liu,[1,2] Feng Liu[1,*]

*[1] Department of Materials Science and Engineering, University of Utah, Salt Lake City, Utah 84112, USA*
*[2] Songshan Lake Materials Laboratory, Dongguan, Guangdong 523808, People's Republic of China*

Phonons as bosons are different from electrons as fermions. Unlike interatomic electron hopping that can be either positive or negative and further tuned by spin-orbit coupling, interatomic spring constant is positive, or the structure of atomic lattices would be dynamically unstable. Surprisingly, we found that topological phonon flat bands (FBs) can manifest either a positive or negative interatomic spring constant that couples the FB-modes of opposite chirality, as exemplified by first-principles calculations of a 2D material of Kagome-BN. To reveal its physical origin, we first establish a fundamental correspondence between a collective lattice-coupling (CLC) variable of two quasi-particle states (e.g., electronic states or phonon modes) of opposite parity in a periodic lattice with band topology. Topological semimetals arise with zero CLC at special *k*-points protected by symmetry; while positive and negative CLC at these *k*-points gives rise to normal and topological insulators, respectively. Then, we show topological FB has a special form of CLC that vanishes at all *k*-points as characterized by its real-space wave function, and multi-atom FB phonon mode can manifest effectively a negative interatomic spring constant. Our findings shed new light on our fundamental understanding of topology and provide a practical design principle for creating artificial bosonic topological states.

[*]fliu@eng.utah.edu

In crystalline solids, quasiparticle transport is mediated by interatomic coupling constants in a periodic lattice, which can be different for fermionic electrons from bosonic phonons. Interatomic electron hopping of different valence orbitals can be either positive or negative; a negative hopping integral ($-t = te^{i\pi}$) implies that the Bloch electrons acquire an additional phase of $\pi$ as they hop over the periodic lattice. Also, the phase of electron hopping can be tuned by spin-orbit coupling (SOC), which is the key ingredient to open a topological band gap. In contrast, the interatomic spring constant, coupling the atomic vibration modes, can only be positive; otherwise, the atomic structure would be dynamically unstable. This fundamental difference is evident from the observation that the sign of electron eigenvalues can be negative or positive, relative to the position of Fermi level, while all the phonon eigen frequencies must be positive. Furthermore, phonon is spinless lacking SOC to tune the phase of spring constant.

Let's further illustrate the difference between interatomic electron hopping parameter and interatomic spring constant, in the context of topological flat band (FB). We take Kagome lattice (Fig. 1(a)) as a prototypical example which is well-known to host FB [1-3]. A positive and negative electron hopping (*t*) leads to a FB of opposite chirality sitting above and below the dispersive Dirac bands [4,5], respectively, as shown in Fig. 1(b) and 1(c). Now we examine a phononic Kagome lattice (Fig. 1(d)), where each lattice site is bonded with four nearest-neighbor (NN) sites. The z-mode phonon dynamical matrix is given in Eq. (1a) (see Methods section of Supplementary Materials (SM) [6]):

$$D\boldsymbol{u} = k^f \begin{pmatrix} 4 & -1 - e^{\frac{i}{2}k_x - \frac{i\sqrt{3}}{2}k_y} & -1 - e^{-\frac{i}{2}k_x - \frac{i\sqrt{3}}{2}k_y} \\ -1 - e^{-\frac{i}{2}k_x + \frac{i\sqrt{3}}{2}k_y} & 4 & -1 - e^{-ik_x} \\ -1 - e^{\frac{i}{2}k_x + \frac{i\sqrt{3}}{2}k_y} & -1 - e^{ik_x} & 4 \end{pmatrix} \begin{pmatrix} u_{A,z} \\ u_{B,z} \\ u_{C,z} \end{pmatrix}, \quad (1a)$$

$$H\boldsymbol{\psi} = t_{pp\pi} \begin{pmatrix} 0 & -1 - e^{\frac{i}{2}k_x - \frac{i\sqrt{3}}{2}k_y} & -1 - e^{-\frac{i}{2}k_x - \frac{i\sqrt{3}}{2}k_y} \\ -1 - e^{-\frac{i}{2}k_x + \frac{i\sqrt{3}}{2}k_y} & 0 & -1 - e^{-ik_x} \\ -1 - e^{\frac{i}{2}k_x + \frac{i\sqrt{3}}{2}k_y} & -1 - e^{ik_x} & 0 \end{pmatrix} \begin{pmatrix} \psi_A \\ \psi_B \\ \psi_C \end{pmatrix}. \quad (1b)$$

In deriving Eq. (1a), without losing generality and for simplicity, we set atomic mass $M_{A,B,C}=1$, lattice constant $a = 1$ and interatomic spring constant $k^f=1$ (superscript *f* is used to distinguish $k^f$ from momentum *k* throughout). Note that while the interatomic spring constant $k^f$ must be positive, the elements of force constant matrix can have different signs and those of dynamic matrix further varies with *k* (see detailed discussion in Methods section of SM [6]).

The calculated phonon bands are shown in Fig. 1(e). One sees that there is a FB, similar to the electronic FB (Fig. 1(b)). For comparison, we give the electronic tight-binding Hamiltonian $H$ for $p_z$-orbitals in a Kagome lattice in Eq. (1b). The two matrices are same except different diagonal elements, so that their eigen spectra appear similar, as shown in Fig. 1(b) and 1(e) for positive $t$ and $k^f$, and Fig. 1(c) and 1(f) for negative $t$ and $k^f$ (to be assumed and discussed below), respectively. This analogy shows that in principle propagation of z-mode vibrations can be qualitatively understood as hopping of electron $p_z$-states in a lattice.

However, there are two fundamental differences to be noticed. First, the diagonal elements of $H$ represent the atomic valence orbital on-site energies, which can be set positive or negative in accordance with eigenvalues relative to the Fermi energy. In contrast, the diagonal elements of $D$ represent the number of NN springs connecting to the atomic site (4 in the present case), and the eigen frequency for the acoustic phonon frequency accounting for translational motion of atoms tends to zero at $\Gamma$. Secondly, the atomic valence orbitals can have $s$, $p$, $d$ and $f$ symmetries whose NN hopping integrals [$t$ in Eq. (1b)] can be either positive or negative. Correspondingly, the position of electronic FB can be either above (Fig. 1(b)) or below (Fig. 1(c)) the Dirac bands for $+t$ or $-t$, respectively [4,5]. In contrast, the atomic vibrational modes have only $p$ symmetry and the interatomic spring constant [$k^f$ in Eq. (1a)] can only be positive. If one artificially assumed a negative $k^f$, one would get a phonon band structure as shown in Fig. 1(f), which has the FB below the Dirac bands, as expected from the analogy with electronic bands, but all the vibration frequencies would be negative (imaginary). This is apparently unphysical. It seems that in a Kagome lattice, phononic FB should always sit above the Dirac bands.

Surprisingly, we found that a phonon FB sits below the Dirac bands, manifesting a negative $k^f$, do exist in real materials such as Kagome-BN (Fig. 2(a)), as shown in Fig. 2(b), from density-functional-theory (DFT) calculations (See Methods in SM [6]). Kagome-BN was previously studied for its electronic Dirac bands and heavy fermion properties [7], while its phonon spectrum was also calculated which we reproduce here. In Fig. 2(b), there are two sets of Kagome phonon FBs sitting under Dirac bands in the optical branches between 25 and 50 THz. Hypothetically, one could fit them separately as if they arose from two Kagome lattices with one atom per lattice site (assuming 1 atomic mass unit (*amu*) for atoms). Then one would obtain an on-site single-atom mode frequency of 30.5 THz and 41.1 THz, respectively, with a negative $k^f$

$= -2882$ N/m.

To reveal its physical origin, we became to realize that first, one has to consider multi-atom cluster-mode vibrations, because the single-atom vibration modes cannot have a negative $k^f$. Secondly, one may expect the negative (or sign of) $k^f$ is related to phonon topology, because it is known that two electronic FBs arising from different sign of $t$ (Fig. 1(b) and 1(c)) have opposite chirality, i.e., they have opposite Chern number if SOC is added and time reversal symmetry is broken [2,5,8]. This points to an analogous correlation between the sign of $k^f$ (Fig. 1(e) and 1(f)) and phonon FB chirality. Therefore, to resolve a negative $k^f$, one needs to further understand the relationship between the multi-atom phonon modes and phonon topology, especially the FB topology. In this Letter, we establish generally a fundamental correspondence between a collective lattice-coupling (CLC) variable of two quasi-particle quantum states of opposite parity in a periodic lattice with band topology, which underlines the mechanism of band inversion between two multi-atom cluster modes of opposite parity. Then, we show that topological FB has a special form of CLC that vanishes at all the $k$-points in the whole Brillouin zone (BZ), which can be characterized by its real-space wave function and topology, and demonstrate that only the multi-atom topological phonon FB can manifest a negative interatomic spring constant.

We will use electrons and phonons as examples to establish the CLC-topology correspondence which is generally applicable to all quasiparticles. Let us begin with a brief review of band topology induced by band inversion using the well-known 1D Su–Schrieffer–Heeger (SSH) model as shown in Fig. 3(a-c). Let $t_1$ ($t_2$) be the intracell (intercell) interatomic hopping as shown in Fig. 3(a). The intracell hopping $t_1$ actually defines an "on-site" level splitting within the unit cell between two electronic states of opposite parity $(|A\rangle \pm |B\rangle)/\sqrt{2}$ (odd-even). The effect of intercell hopping $t_2$ is to change the level splitting at momentum $k$ (i.e., band dispersion) as these two states hop in the lattice periodically from cell to cell. One notices that at the BZ boundary ($ka = \pm\pi$, $a$ is lattice constant), the level spitting becomes $t_1 - t_2$. Then, if $t_1 > t_2$, the order of level splitting and hence the phase of the two quantum states is unchanged as they hop over the lattice, so that the system is topological trivial (Fig. 3(b)). If $t_1 < t_2$, however, the order of level splitting is reversed (band inversion) and the system becomes topological nontrivial (Fig. 3(c)). This observation invoked us to introduce a new variable of CLC between A and B sublattices for electron hopping, defined as $T_{A\to B} = t_1 + t_2 e^{-ika}$, which accounts for both interatomic hopping strength and structure factor, to capture the phase evolution of the two states $(|A\rangle \pm |B\rangle)/\sqrt{2}$ over the BZ. At the BZ boundary, $T_{A\to B} = t_1 - t_2$ whose sign defines the band topology: a positive (negative) $T_{A\to B}$ corresponds to a trivial (nontrivial) band topology, as shown in Fig. 3(b) (Fig. 3(c)).

An analog of phononic SSH model can be constructed, as shown in Fig. 3(d-f). Similarly, we define a CLC of force constants coupling sublattice A and B, $K^f_{A\to B} = k_1^f + k_2^f e^{-ika}$, which accounts for interatomic spring constant and structure factor, as well as directional cosines for in-plane modes (see Methods in SM [6]). At the BZ boundary ($X$ point), $K^f_{A\to B} = k_1^f - k_2^f$. Then, if $K^f_{A\to B} > 0$, the phonon bands are topological trivial (Fig. 3(e)); if $K^f_{A\to B} < 0$, they are nontrivial (Fig. 3(f)). To better understand this, we examine the eigen modes at $\Gamma$ and $X$. At $\Gamma$ (Fig. 3(g)), the eigen modes have one acoustic branch at bottom (even parity) and another optical branch on top (odd parity). This order of modes is unchanged at $X$ when $K^f_{A\to B} > 0$ (Fig. 3(h)), but reversed when $K^f_{A\to B} < 0$ (Fig. 3(i)). Interestingly, for $K^f_{A\to B} < 0$, i.e., $k_1^f < k_2^f$, the diatomic phonon modes can be viewed shifted by half a unit cell, as shown by the dashed ovals in Fig. 3(i) shifted from Fig. 3(g), to become out of phase in the neighboring unit cells. Then the optical mode is seen to have two atoms in one unit cell breathing out while the two atoms in the neighboring cell breathing in (upper panel of Fig. 3(i)), as if there is no restoring force. Such behavior of optical mode switching is previously known; here, we use it to explain the physical underpinning of a negative CLC of force constants coupling the two-atom cluster modes, leading to "mode inversion" as the modes propagate over the lattice and hence a nontrivial phonon band topology. We note that the phonon topology via mode inversion corresponds to the sign of CLC of force constants as we define here, not the sign of individual element of force constant or dynamic matrix.

The above correspondence between the CLC variable of two quantum states and topology is general and equally applicable to 2D lattices. For example, consider the graphene lattice, there are two $\pi$ and $\pi^*$ electronic states per unit cell resulting from linear combination of $p_z$-orbitals on A and B sublattices ($(|p_{z,A}\rangle \pm |p_{z,B}\rangle)/\sqrt{2}$), whose CLC of electron hopping is calculated as $T_{A\to B} = t_1 + t_2(e^{-ika_1} + e^{-ika_2})$, where the first ($t_1$) term is the intracell and the second ($t_2$) term is the intercell NN hopping, respectively, and $a_1$ ($a_2$) are lattice vectors (for details, see Fig. S1 in SM [6]). By $C_3$ rotation symmetry ($t_1 = t_2$), one finds $T_{A\to B} = 0$ at the corners of BZ ($K$ and $K'$ points). Thus, $T_{A\to B} = 0$ corresponds to a topological state of Dirac semimetal protected by symmetry. Following the seminal work by Kane-Mele [9], if one breaks inversion symmetry by adding an on-site energy difference between A and B sublattice, the system opens a trivial gap becoming a normal insulator, which can be also understood as it effectively increases $t_1$ making the CLC positive at ($K$, $K$'). On the other hand, if one adds SOC, it effectively adds a phase to the $t_2$ term making the CLC negative at ($K$, $K$'), so that the system opens a nontrivial gap becoming a topological insulator. Similarly, a CLC of force constants between A and B sublattices can be defined for graphene phonons.

For electrons, the CLC-topology correspondence established above applies also to cases of single atom per unit cell, because atomic valence orbitals have different (*s*, *p*, *d* ...) parities and interatomic electron hopping can be positive or negative. One can reformulate the seminal work by Bernevig-Hughes-Zhang of band inversion model in a rectangle [10], triangle [11] or square lattice by defining a CLC of electron hopping $T_{s\to p}$ between the *s*- and *p*-orbitals in presence of the *p*-orbital-originated SOC (for details see Fig. S3 in SM [6]). For phonons, however, the correspondence principle applies only to multi-atom per unit cell, because single atomic vibrations have only *p*-symmetry and interatomic spring constant is always positive so that a CLC cannot be defined. This means phonons are always topological trivial in all materials with single atom per unit cell.

Now we come back to analyze phonon FB in terms of both the CLC of multi-atom phonon modes and the interatomic spring constant. As shown above, a topological phonon Dirac state corresponds to a zero CLC at the Dirac *k*-points. It presents local phase cancellation of the Bloch wave functions of two phonon modes of opposite parity to form a Berry flux center at the Dirac *k*-points. This correspondence applies also to nodal-line semimetals where the CLC vanishes at *k*-points along a line in BZ (see Fig. S2 and section II in SM[6]). Differently, a topological phonon FB, such as in Kagome lattice or more generally in line-graph lattices [12], arises from destructive interference of Bloch states. It is dispersionless with zero kinetic energy (zero group velocity), because the phases of FB-phonon Bloch wave functions cancel out with each other globally at all the *k*-points in the BZ, independent of the magnitude of interatomic spring constant $k^f$. This means that implicitly the phonon FB-mode has a special form of CLC, not expressible as an explicit function of $k^f$ and momentum *k*, different from the one defined above for local phase cancellation and band inversion. Instead, it can be inferred from the real-space FB-phonon wave function, the so-called compact localized state [12,13], as shown in Fig. 2(c), calculated from Fig. 1(e). It has six nodal points at the vertices of a hexagon with red and blue circles denoting the positive (up) and negative (down) displacement along z-axis, respectively. Thus, the FB-mode is completely localized in real space, as reflected by cancellation of pair-wise restoring forces acting outside the hexagon (red and blue dashed arrows in Fig. 2(c)). This signifies a destructive quantum interference of phonon wave function induced by Kagome lattice symmetry, in analogy to compact localized state of electronic FB [12,13]. Effectively, the compact localized state in Fig. 2(c) can be seen as a flux center, where the phase of wave function evolves periodically around a "loop", indicating the FB has a real-space topology [14] without SOC or band inversion. Alternatively, the singular band touching point at $\Gamma$ (Fig. 1(e)) can be viewed as a Berry flux center in $k$-space in analogy to Dirac point [12-13].

However, a topological phonon FB will only manifest a positive $k^f$ in a Kagome lattice with one atom per lattice site, as shown earlier in Fig. 1(d). Therefore, one must consider the case of multi atoms per Kagome lattice site, to derive a FB with a negative $k^f$. For example, we have found that the out-of-plane vibrational modes of di-atomic and tri-atomic Kagome lattices can have FBs manifesting effectively a negative interatomic (i.e., inter-Kagome-lattice-site) spring constant (see Fig. S4 and Section IV in SM [6]). Here, we focus on explaining the surprising results of Kagome-BN in Fig. 2(b) by considering the in-plane vibration modes from a four-atom cluster (two B and two N atoms) on each Kagome lattice site. As shown in Fig. 2(a), there are three different springs: N-N bond ($k_{11}^f$), B-N bond ($k_{12}^f$) and B-B bond ($k_{22}^f$), and two atoms of different mass: B (11 *amu*) and N (14 *amu*). From the calculated phonon spectra which contains 24 bands (see Fig. S7 in SM [6]), we found two sets of Kagome bands in the frequency range of ~25-50 THz, as shown in Fig. 2(d), which agree nicely with the DFT results featured with two FBs both manifesting a negative interatomic spring constant. It is worth noting that the model calculations are done by setting $k_{11}^f$ = 8000 N/m, $k_{12}^f$ = 5600 N/m, $k_{22}^f$ = 2500 N/m, which are perfectly consistent with the order of bond lengths (N-N bond: 1.02 Å, B-N bond: 1.36 Å, B-B bond: 1.70 Å) or strength. Also, different atomic mass of B and N are used which affects the results quantitatively but not qualitatively.

Finally, we comment on some general implications of the intriguing negative interatomic spring constant manifested by multi-atom topological phonon FB and the newly established CLC-topology correspondence. It provides a unified view (a "descriptor") for three common classes of topological materials: topological semimetal, insulator and FB materials. First, topological semimetals correspond to a zero CLC at special *k*-points. This can be generally achieved for both fermions and bosons by symmetry operations in a crystal. Most topological phonons found in real materials belong to this category, such as phononic Dirac/Weyl point [15-19] and nodal-line states [20-25]. Secondly, topological insulators, including high-order topological insulators (see Fig. S5-S6 and section V in SM [6]), correspond to a negative CLC at the special *k*-points. Without SOC, this is achievable for both electrons and phonons through alternation of single and double bonds between the same atoms to modulate bond strength, such as polyacetylene to realize the SSH model [26] and graphdiyne to realize the Kekulé model of high-order topological-insulator [27]. On the other hand, starting from a topological semimetal or a narrow-gap semiconductor with *s*- and *p*-state band edges, most topological insulators are achieved by adding SOC to turn the CLC negative. In this sense, all electronic materials could be topological with SOC, provided it has the right lattice constant and position of Fermi level [28]. However, it is much harder for phonons lacking SOC to attain a negative CLC. Another fundamental difference is that the electronic bands of single atom per unit cell can be topological because a CLC, such as between atomic *s*- and *p*-valence, can be defined; while the phonon bands of single atom per unit cell are always trivial because a CLC cannot be defined. For these reasons, so far phonon topological insulators have been studied by models *implicitly* assuming a negative CLC of force constants [29-32], but rarely in real materials. On the other hand, the correspondence principle we establish for phonons can be readily applied to designing bosonic topological states in artificial, such as photonic, acoustic and mechanical-wave systems [32-36], where CLC can be manipulated by design. Especially, FB corresponds to a zero CLC at all *k*-points. Given that a phonon FB can manifest a negative interatomic spring constant, more generally, bosonic FBs of opposite chirality can be manufactured in artificial systems by designing positive versus negative "interatomic" coupling.

## Acknowledgements

We thank T.L. Feng for helpful discussions. B.X. and F.L. are supported by U.S. DOE-BES (Grant No. DE-FG02-04ER46148). H.L. acknowledge financial support from China Postdoctoral Science Foundation (Grant No. 2021M700163).

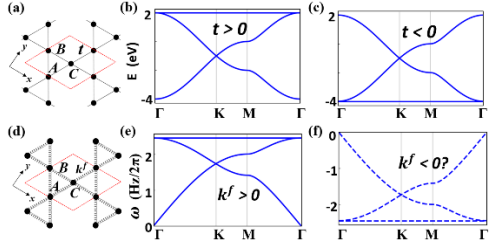

Fig. 1. Illustration of electron (phonon) FB with positive vs. negative interatomic hopping (spring constant). (a) The electronic Kagome lattice with hopping $t$. (c) The electron band structure for $t > 0$. (c) The electron band structure for $t < 0$. (d) Spring-mass model for Kagome lattice with out-of-plane z-mode spring constant $k^f$. (e) The phonon spectra for $k^f > 0$. (f) The phonon spectra for $k^f < 0$.

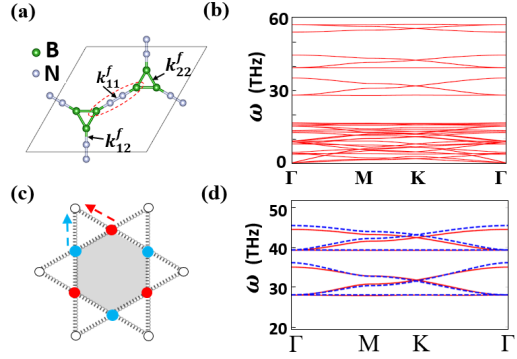

Fig. 2. Demonstration of negative interatomic spring constant manifested by phonon FB in Kagome-BN. (a) The structure of 2D Kagome-BN, B and N atoms are denoted by green and grey sphere, respectively. (b) The DFT calculated phonon spectra for Kagome-BN. (c) The compact localized state of Kagome FB. (d)The comparison between model (blue dashed lines) and DFT (red solid line) results for two subsets of Kagome bands manifesting a negative interatomic spring constant.

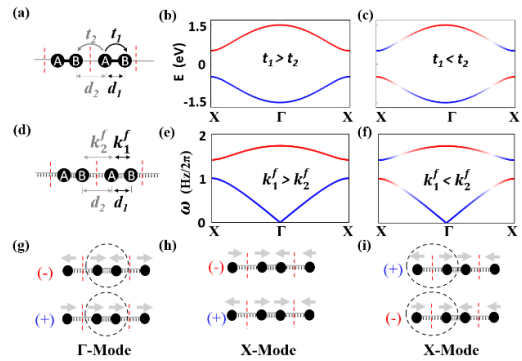

Fig. 3. Illustration of electron (phonon) band topology in correspondence with positive vs. negative CLC in SSH model. (a) Electronic SSH model. The red dashed lines mark the unit cell. (b) Band structure for $T_{A-B} > 0$ $(t_1 > t_2)$. The even (odd) parity of band is marked by red (blue) color. (c) Band structure for $T_{A-B} < 0$ $(t_1 < t_2)$. (d) Phononic SSH model. (e) The phonon spectra for $K^f_{A-B} > 0$ $(k^f_1 > k^f_2)$. (f) The phonon spectra for $K^f_{A-B} < 0$ $(k^f_1 < k^f_2)$. (g) The eigenmodes at $\Gamma$. (h) The eigenmodes at $X$ for $K^f_{A-B} > 0$. (i) The eigenmodes at $X$ for $K^f_{A-B} < 0$.

Supplemental Material for

Negative Interatomic Spring Constant Manifested by Topological Phonon Flat Band

Bowen Xia,[1] Hang Liu,[1,2] Feng Liu[1,*]

[1] *Department of Materials Science and Engineering, University of Utah, Salt Lake City, Utah 84112, USA*
[2] *Songshan Lake Materials Laboratory, Dongguan, Guangdong 523808, People's Republic of China*

# I. Methods

## I.a. Spring-mass model

Within the harmonic-oscillator approximation (spring-mass model), the eigen equation of phonon dynamical matrix ($D$) is written as:

$$(\omega^2 - D)u = 0, \quad (S1)$$

where $\omega$ is the phonon frequency, $\boldsymbol{u} = (u_{1,x}, u_{1,y}, u_{1,z}, u_{2,x} \ldots u_{n,\alpha})^\dagger$ is the atomic displacement. $D_{\alpha\beta}^{mn} = \sum_R \frac{1}{\sqrt{M_m M_n}}(-\Phi_{\alpha\beta}^{mn}(R))e^{i\boldsymbol{k}\cdot(\boldsymbol{r}_n - \boldsymbol{r}_m + R)}$, where $R, M$ are the lattice vector and mass; $m, n$ are the atom indices. $\Phi_{\alpha\beta}^{mn}(R) = k f_{\alpha\beta}^{mn}(R) S_{\alpha\beta}^{mn}(R)$, and $\alpha, \beta$ denote the directions of displacement. We note that $k f_{\alpha\beta}^{mn}$ is the interatomic spring constant as physically defined in the spring-mass model, which can only take *positive* values. But the elements of force constant matrix $\Phi_{\alpha\beta}^{mn}$ can have different signs due to the directional cosine factor of $S_{\alpha\beta}^{mn}$. $\Phi_{\alpha\beta}^{mn}$ can be generally calculated from the second derivative of potential energy, using first-principles or force-field methods, with respect to atomic positions, where the factor $S_{\alpha\beta}^{mn}$ is implicitly included leading to different signs. The elements of dynamics matrix $D_{\alpha\beta}^{mn}$ will further vary and change signs with momentum $k$ due to the structure factor of a periodic lattice. Each atom has three basic phonon modes, corresponding to vibrations in the $x$, $y$, and $z$ directions. In general, atomic displacement causes bond stretching and bending with a directional cosine factor of $S_{\alpha\beta}^{mn} = \cos(\theta_\alpha)\cos(\theta_\beta)$ and $S_{\alpha\beta}^{mn} = \sin(\theta_\alpha)\sin(\theta_\beta)$, respectively, where $\theta_{\alpha,\beta}$ is angle between the displacement vector and bond direction. In planar 2D lattices, out-of-plane phonon mode ($z$-mode) is decoupled from the in-plane modes ($x,y$-modes) and the restoring force for $z$-mode has only the bond-bending component. Thus, $\Phi_{zz}^{mn} = kf_{bending}^{mn}$ for the $z$-mode, and $\Phi_{\alpha\beta}^{mn} = kf_{stretching}^{mn}\cos(\theta_\alpha)\cos(\theta_\beta) + kf_{bending}^{mn}\sin(\theta_\alpha)\sin(\theta_\beta)$ for the in-plane modes. Usually, the spring constant for bond-stretching is orders of magnitude larger than that of bond-bending [1]. Therefore, without losing generality and for simplicity, we neglect the in-plane bending mode in our model analysis.

*fliu@eng.utah.edu

I.b. Density-functional theory (DFT) calculation

The phonon calculations for Kagome-BN were performed using Vienna ab initio simulation package (VASP) [2] within the framework of DFT. The Perdew-Burke-Ernzerhof functional within the generalized gradient approximation was adopted [3-5]. The cutoff energy is set to 400 eV and a 3*3*1 Monkhorst-Pack k-points grid is used to sample the Brillion zone [6]. The structure was fully relaxed until the forces on each atom are smaller than 0.001 eV/ Å. And a 15 Å thick vacuum layer was used to ensure the decoupling between neighboring slabs. The phonon spectra are calculated using PHONOPY package [7] based on the finite displacement method using a 3*3*1 supercell.

I.c. Tight-binding calculation

Tight-binding models are constructed using Slater-Koster parameters [8] to explore the correspondence between topology and collective lattice coupling (CLC) of electron hopping in the $s, p_z$-Kagome lattice, 1D $s, p_z$-SSH model, $s, p_z$-hexagonal lattice, $sp^2$-square lattice, $s, p_z$-Kekulé lattice.

## II. Correspondence between the CLC variable of two quantum states and their topology in graphene lattice: topological phase transition

As shown in Fig. S1(a), in graphene lattice, each sublattice A site can hop to three nearest-neighbor B sites, including one intracell ($t_1$) and two intercell ($t_2$) hopping. The CLC of electron hopping can be defined as $T_{A \to B} = t_1 + t_2(e^{-ika_1} + e^{-ika_2})$, which is effectively the coupling of two sublattices A and B of a bipartite lattice like graphene. By $C_3$ rotation symmetry ($t_1 = t_2$), one finds that $T_{A \to B} = t_1(1 + e^{-ika_1} + e^{-ika_2}) = 0$ at corners of Brillouin zone (BZ) (K and K' points) which corresponds to a topological semimetal protected by symmetry as shown in Fig. S1(b) (Dirac cone at K point). The result of $T_{A \to B} = 0$ at the Dirac points represents a condition satisfied for local phase cancellation of Bloch wave function, leading to formation of Berry flux center at K and K'. If one breaks the inversion symmetry by adding an on-site energy difference between A and B sublattice, for example adding 0.5 eV on-site energy to A sublattice as shown in Fig. S1(c, d), the system opens a trivial gap, becoming a normal insulator, which can be now understood as it effectively increases $t_1$ making $T_{A \to B}$ positive at (K, K'). On the other hand, if one adds spin-orbital coupling (SOC, λ = 0.5 eV) as shown in Fig. S1(e, f), it effectively adds a phase to the $t_2$ term making $T_{A \to B}$ negative at (K, K'), so that the system opens a nontrivial gap becoming a topological insulator.

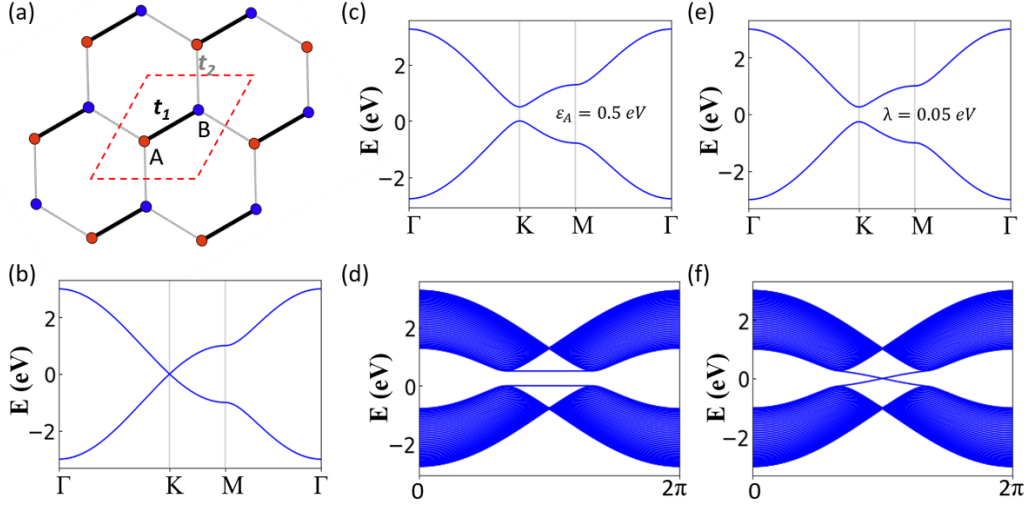

Fig. S1. Illustration of the CLC-topology correspondence in graphene lattice. (a) Graphene lattice. Intra- and inter-cell hopping is marked as $t_1$ and $t_2$, respectively. Sublattice A(B) is denoted by red (blue) dots. (b) The band structure for $t_1 = t_2$. (c-d) The band structure and edge states for $t_1 = t_2$ and $\varepsilon_A = 0.5$ eV, $\varepsilon_B = 0$ eV. (e-f) The band structure and edge states for $t_1 = t_2$, $\varepsilon_A = \varepsilon_B = 0$ and $\lambda = 0.5$ eV.

## II. Correspondence between the CLC variable of two quantum states and their topology in checkerboard lattice: nodal-line semimetal

Here we show that topological nodal-line semimetal corresponds to zero CLC at special lines of $k$-points in BZ, using the checkerboard phonon lattice model as an example. For simplicity, we consider the z-mode. As shown in Fig. S2(a), each sublattice A site couples to four nearest-neighbor B sites, including one intracell ($k_1^f$) and three intercell ($k_2^f$) interatomic spring constants. We define a CLC of force constants as $K_{A \to B} = k_1^f + k_2^f(e^{-i k_x} + e^{-i k_y} + e^{-i(k_x + k_y)})$. Then, if $k_1^f > k_2^f$, $K_{A \to B}$ is positive in the whole BZ, the system is a trivial insulator as shown in Fig. S2(b-c). If by symmetry, $k_1^f = k_2^f = k^f$, $K_{B \to A} = k^f(1 + e^{-i k_x} + e^{-i k_y} + e^{-i(k_x + k_y)})$, which will vanish along the line X-M: $K_{B \to A}(k_{x,y} = \pm\pi) = 0$, corresponding to a topological nodal-line semimetal as shown in Fig. S2(d).

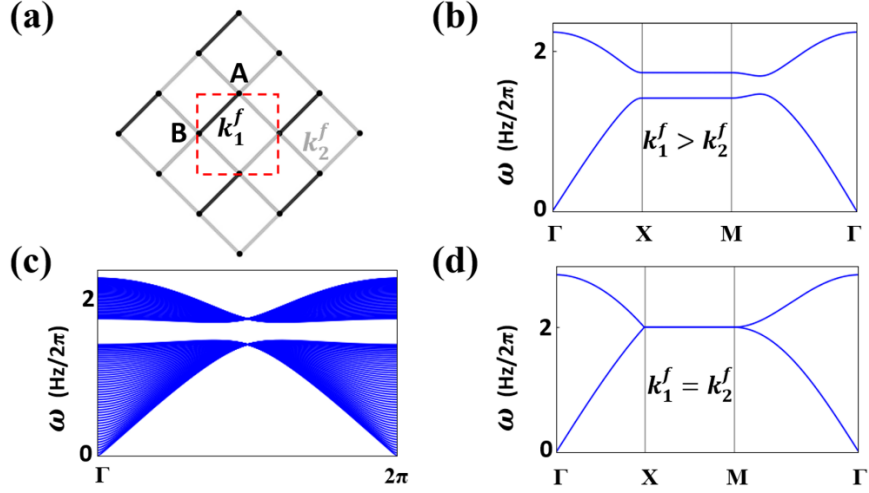

Fig. S2. Illustration of the CLC-topology correspondence in phonon checkerboard lattice for a nodal-line semimetal. (a) Checkerboard lattice. Intra- and inter-cell spring constant is marked in black ($k_1^f$) and grey ($k_2^f$), respectively. (b) The phonon spectrum for $k_1^f > k_2^f$. (c) The lack of edge states for $k_1^f > k_2^f$. (d) The phonon spectrum for $k_1^f = k_2^f$, showing a nodal line from $X$ to $M$.

## III. Electronic $sp^2$ model in a square lattice

In this section, we will demonstrate that our established CLC-topology correspondence can be applied to cases of single atom per unit cell for electronic states. Here, we take a $sp^2$ square lattice as an example, as shown in Fig. S3(a). The Hamiltonian can be written as Eq. S2 using ($s$, $p_x + ip_y$, $p_x - ip_y$) basis:

$$H = \begin{pmatrix} \varepsilon_s + 2t_{ss\sigma}(\cos(k_x) + \cos(k_y)) & \sqrt{2}t_{sp\sigma}(i\sin(k_x) - \sin(k_y)) & \sqrt{2}t_{sp\sigma}(i\sin(k_x) + \sin(k_y)) \\ \sqrt{2}t_{sp\sigma}(-i\sin(k_x) - \sin(k_y)) & \varepsilon_p + \lambda + (t_{pp\pi} + t_{pp\sigma})(\cos(k_x) + \cos(k_y)) & -((t_{pp\pi} - t_{pp\sigma})(\cos(k_x) - \cos(k_y))) \\ \sqrt{2}t_{sp\sigma}(-i\sin(k_x) + \sin(k_y)) & -((t_{pp\pi} - t_{pp\sigma})(\cos(k_x) - \cos(k_y))) & \varepsilon_p - \lambda + (t_{pp\pi} + t_{pp\sigma})(\cos(k_x) + \cos(k_y)) \end{pmatrix} \quad (S2)$$

Without SOC ($\lambda = 0$), two $p$-bands are degenerate at $\Gamma$ as shown in Fig. S3(b) ($t_{ss\sigma} = t_{sp\sigma} = 1$, $t_{pp\sigma} = -1$, $t_{pp\pi} = -0.1$, $\varepsilon_s = -4.5$ and $\varepsilon_p = 2$). If one gradually increases SOC strength $\lambda$, the $p$-bands degeneracy is lifted. And if the SOC is strong enough, one of the $p$-bands will further be inverted with $s$-band at $\Gamma$ and a topological gap opens as shown in Fig. S3(c-d) with $s$-$p$ band inversion.

To show that our newly established correspondence principle can be applied to explain the above topological phase transition, we first expand Eq. S2 to the 1$^{st}$ order around $\Gamma$ as shown in Eq. S3:

$$H(\Gamma) = \begin{pmatrix} \varepsilon_s + 4t_{ss\sigma} & -\sqrt{2}t_{sp\sigma}\,k_y + i\sqrt{2}\,t_{sp\sigma}\,k_x & \sqrt{2}t_{sp\sigma}\,k_y + i\sqrt{2}\,t_{sp\sigma}\,k_x \\ -\sqrt{2}t_{sp\sigma}\,k_y - i\sqrt{2}\,t_{sp\sigma}\,k_x & \varepsilon_p + \lambda + 2(t_{pp\pi} + t_{pp\sigma}) & 0 \\ \sqrt{2}t_{sp\sigma}\,k_y - i\sqrt{2}\,t_{sp\sigma}\,k_x & 0 & \varepsilon_p - \lambda + 2(t_{pp\pi} + t_{pp\sigma}) \end{pmatrix} \quad (S3)$$

Then, the CLC of electron hopping between $s$ and $(p_x - ip_y)$ around $\Gamma$ can be obtained by the on-site energy difference (effectively the intracell hopping) between the two states plus their nearest-neighbor lattice hopping (the intercell hopping), $T_{s \to p_x - ip_y}(\Gamma) = \varepsilon_p - \lambda + 2(t_{pp\sigma} + t_{pp\pi}) - \varepsilon_s - 4t_{ss\sigma} + \sqrt{2}t_{sp\sigma}\,k_y + i\sqrt{2}\,t_{sp\sigma}\,k_x$, which reduces to $0.3 - \lambda$ for $t_{ss\sigma} = t_{sp\sigma} = 1$, $t_{pp\sigma} = -1$, $t_{pp\pi} = -0.1$, $\varepsilon_s = -4.5$, $\varepsilon_p = 2$ and $k_x = k_y = 0$. Thus, if $\lambda < 0.3$, $T_{s \to p_x - ip_y} > 0$, the system is trivial as shown in Fig. S3(b); if $\lambda = 0.3$, $T_{s \to p_x - ip_y} = 0$, the system is a topological semimetal as shown in Fig. S3(c); if $\lambda > 0.3$, $T_{s \to p_x - ip_y} < 0$, the system opens a topological gap at $\Gamma$ point as shown in Fig. S3(d).

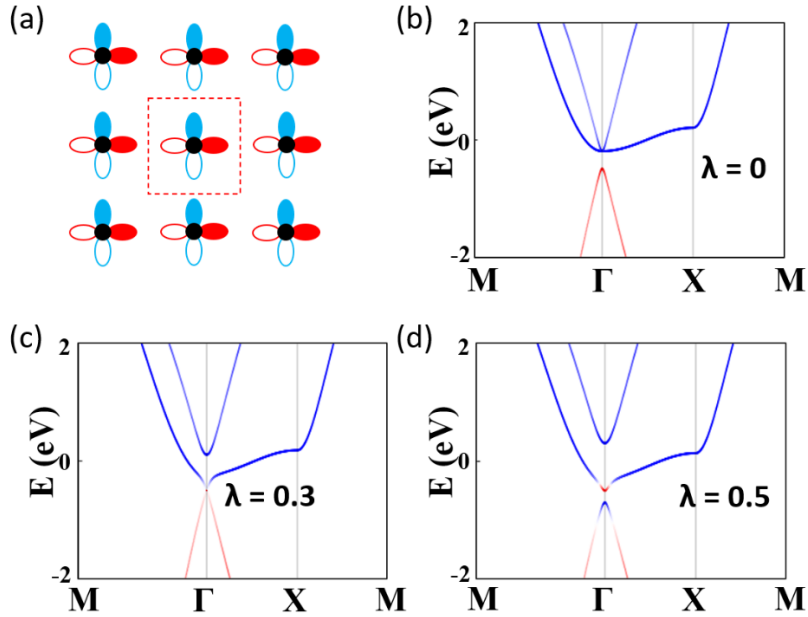

Fig. S3. Illustration of CLC-topology correspondence for $sp^2$ states in a square lattice with single atom per unit cell. (a) Electronic $sp^2$ square lattice. $s$, $p_x$, $p_y$ orbitals are colored in black, red and blue, respectively. (b) The band

structure for λ = 0. The *s*- and *p*-orbital contribution are denoted by red and blue color, respectively. (c) The band structure for λ = 0.3. (d) The band structure for λ = 0.5.

## IV. Out-of-plane phonon bands of di- and tri-atomic Kagome lattices

In this section, we use out-of-plane mode of di- and tri-atomic Kagome lattice as examples to show that FB which corresponds to zero CLC at all *k*-points can manifest either a positive or negative interatomic spring constant that couples the FB-modes of opposite chirality. Let's assume the two (three) atoms on each lattice site are coupled with a spring constant $k_1^f$, and they are coupled with each other on different lattice sites with $k_2^f$, as shown in Fig. S4(a, d). There are two (three) sets of Kagome bands, as the system can be thought of having several Kagome sublattices. Interestingly, one can find that if $k_1^f > \frac{3}{2}k_2^f$, all the subsets of Kagome bands have the FB above the Dirac bands, as shown in Fig. S4(b, e), indicating the FB-mode of a superatom (i.e. a cluster contains two or three atoms as marked by black dashed oval in Fig. S4(a) and (d)) is coupled with a positive interatomic spring constant as indicated by Fig. 1(e). On the other hand, if $k_1^f < \frac{3}{2}k_2^f$, the FB in the lower set of Kagome bands will move up to the upper set sitting below the Dirac bands, as shown by the red FB in Fig. S4(c, f). Consequently, this FB will now manifest a negative interatomic spring constant as indicated by Fig.1(f).

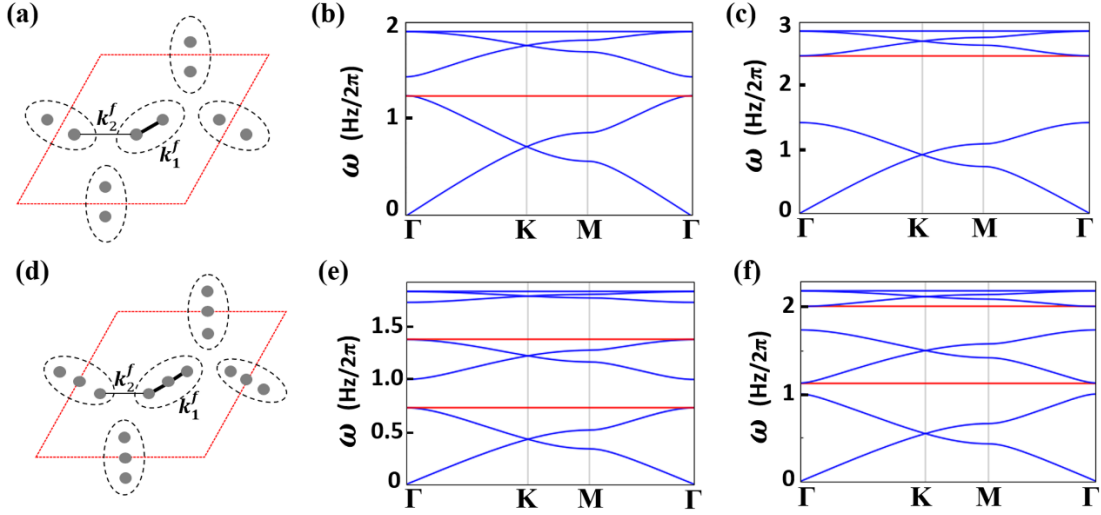

Fig. S4. Di- and tri-atomic Kagome lattice model. (a) Di-atomic Kagome lattice. The black dashed ovals mark the Kagome-lattice-site. (b) The phonon spectrum for $k_1^f > \frac{3}{2} k_2^f$. The red line marks the FB manifesting positive interatomic spring constant between the superatoms on each Kagome-lattice-site. (c) The phonon spectrum for $k_1^f < \frac{3}{2} k_2^f$. The red line marks the FB manifesting negative interatomic spring constant between superatoms on each Kagome-lattice-site. (d) Tri-atomic Kagome lattice. The black dashed ovals mark the Kagome-lattice-site. (e) The phonon spectrum for $k_1^f > \frac{3}{2} k_2^f$. The red lines mark the set of Kagome bands manifesting positive interatomic spring constant between superatoms on each Kagome-lattice-site. (f) The phonon spectrum for $k_1^f < \frac{3}{2} k_2^f$. The red lines mark the set of Kagome bands manifesting negative interatomic spring constant between superatoms on each Kagome-lattice-site.

## V. High-order topological states in Kekulé lattice

In this section, we will elaborate on an alternative understanding of 2D high-order topological states (HOTIs) based on our newly established CLC-topology correspondence, using the Kekulé lattice [9] as an example which is known to host HOTI corner states for both electrons and phonons [10-12]. Kekulé lattice can be considered as deformed from hexagonal lattice as shown in Fig. S5(a), with the intracell and intercell hopping $t_1$ and $t_2$, respectively. Topological corner states will emerge in a finite sample when $t_2 > t_1$. As an alternative understanding of 2D HOTI, namely a 1D first-order TI, one can consider a 1D sublattice of the

2D lattice, by cutting a ribbon along lattice vector $\vec{a}$ direction as shown in Fig. S5(b). By tuning the intracell $t_1$ and intercell $t_2$ hopping, there will be a gap closing and re-opening process near Fermi level as shown in Fig. S5(c-e). To show the band inversion (topology) of the gap, we plot the $\Gamma$-point wave function of the two bands near Fermi level as shown by the inset of Fig. S5(c, e). One can easily see that when $t_2 > t_1$, the wave function of two bands is inverted, indicating a topological gap, similar with 1D SSH model. We numerically calculated the relationship between the gap at $\Gamma$ and $(t_1 - t_2)$ as shown in Fig. S5(f). One can see that the gap is linear to the absolute value of $(t_1 - t_2)$. Thus, the CLC of electron hopping (gap) between the two states at $\Gamma$ around Fermi level can be expressed as $T = 2(t_1 - t_2)$.

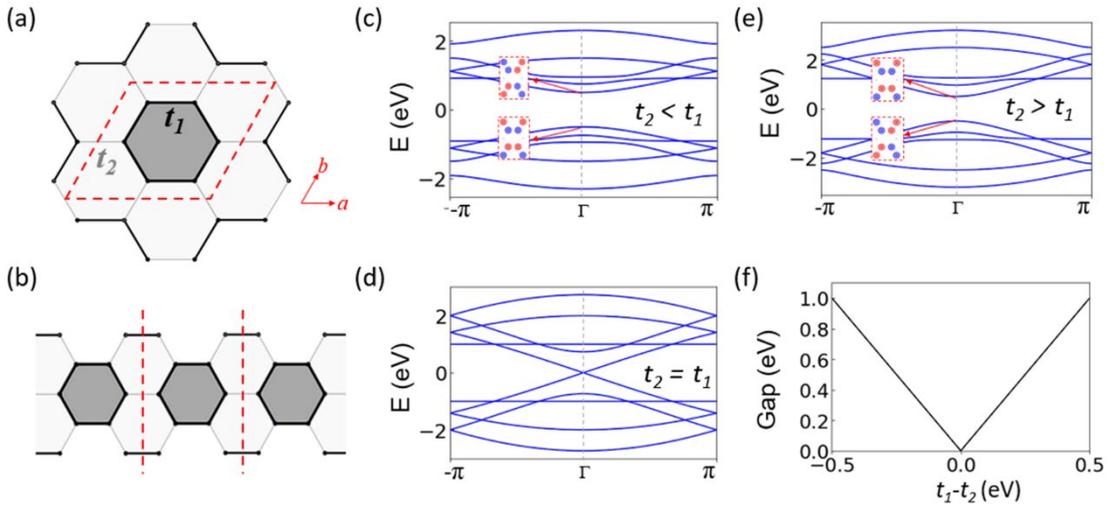

Fig. S5. Illustration of high-order topological states in Kekulé lattice. (a) Kekulé Lattice. (b) 1D armchair nanoribbon of Kekulé lattice. (c) The band structure of nanoribbon when $t_2 < t_1$, the insets show the $\Gamma$-point wave function. (d) The band structure of nanoribbon when $t_2 = t_1$. (e) The band structure of nanoribbon when $t_2 > t_1$, the insets show the $\Gamma$ point wave function. (f) The plot of gap as a function of $(t_1 - t_2)$.

Next, to show the correspondence between CLC of hopping and topology, one can cut a finite-length armchair ribbon with strong bond ($t_2$) broken as shown in Fig. S6(a). For $T > 0$ ($t_2 < t_1$), the energy spectrum is shown in Fig. S6(b) which shows a trivial gap without corner states in the gap. For $T < 0$ ($t_2 > t_1$), there are two topological in-gap states as shown in the red dashed box in Fig. S6(c), which are similar to 1D SSH model with strong bond ($t_2$) broken in

Fig. 2(c). We plot these two states in real-space which are localized at two ends of the ribbon as shown by red dots in Fig. S6(a). Thus, the six corner states of a hexagon flake of Kekulé lattice in Fig. S6(d) can be understood as three pairs of in-gap "end" states of 1D armchair ribbon. Also similar with 1D SSH model (Fig. 2(b)), assuming $t_2 > t_1$, if one makes a finite-length ribbon with weak bond ($t_1$) broken as shown in Fig. S6(e), there will be no topological in-gap states as shown in Fig. S6(f).

On the other hand, for 1D zigzag ribbon as shown in Fig. S6(g), it is a metallic ribbon with many trivial dangling-bond edge states around the zero energy (Fermi level) as shown in Fig. S6(h), similar to zigzag graphene ribbon [13-15]. Consequently, the 1D zigzag ribbon is trivial without in-gap topological states. To further confirm this, one can plot the states near the Fermi level (marked by dashed red box) in real-space as shown Fig. S6(h) which are all localized at zigzag edge rather than at two ends ("corners"). In this way, only two corners of a rhombus sample having corner states can be understood as topological "end" states of one topological nontrivial armchair chain, while the other two corners without corner states are ends of one topological trivial zigzag chain, as shown in Fig. S6(i). In general, one can apply the newly established correspondence principle to identify a $d$-dimensional HOTI by defining a CLC in the ($d$-1)-dimension for electrons or phonons.

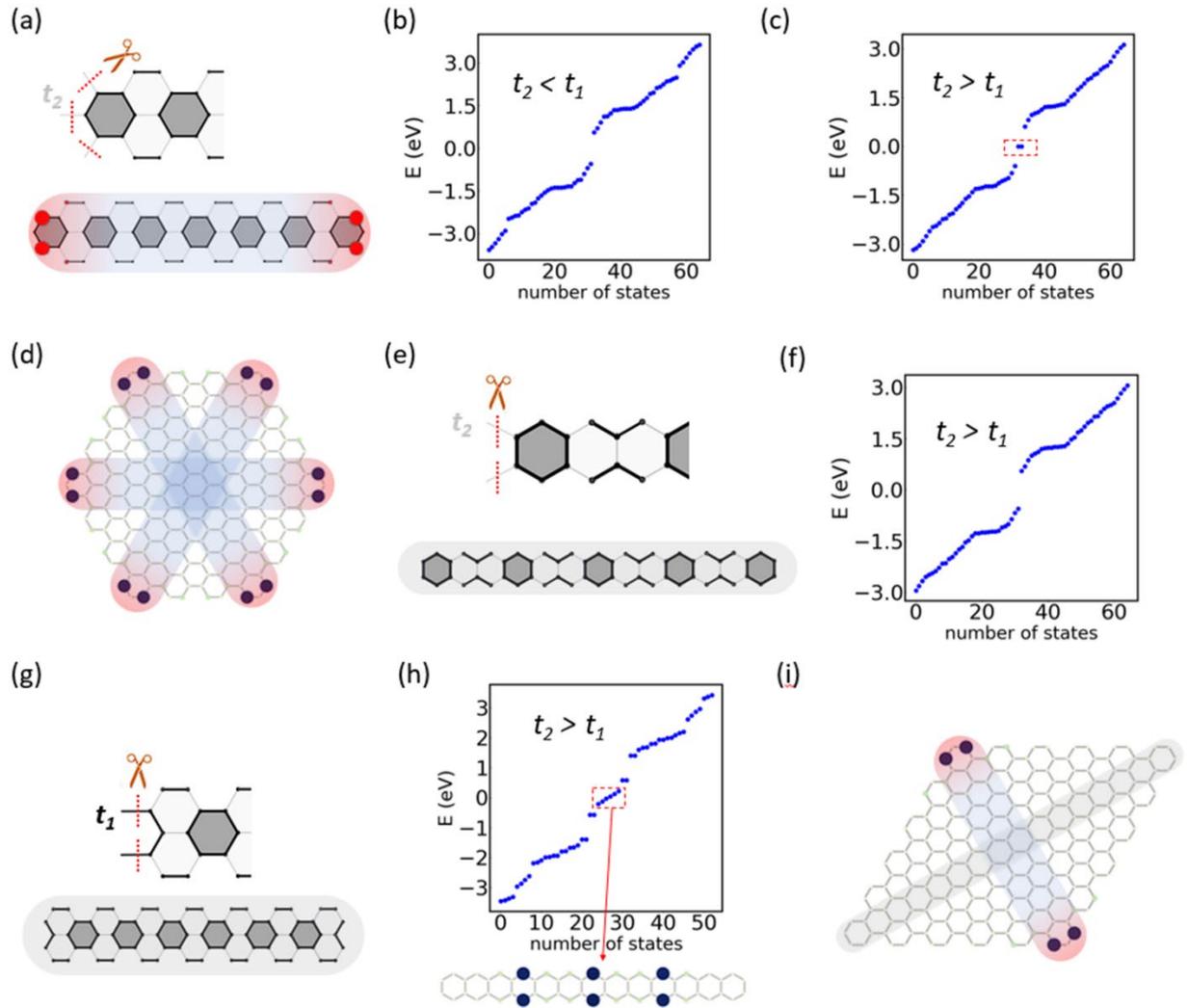

Fig. S6. Illustration of topological "end" state in finite-size Kekulé ribbon. (a) The upper figure shows the way to cut an armchair ribbon with strong bond ($t_2$) broken at the ends. The lower figure shows the finite-size armchair ribbon with the in-gap topological states at the ends. (b) The energy spectrum for T > 0 ($t_2 < t_1$). (c) The energy spectrum for T < 0 ($t_2 > t_1$). The red dashed box marks the in-gap topological "end" states. (d) The corner states of a hexagonal sample. Six corner states can be seen as three sets of boundary (end) states of 1D TI as marked by three colored bands. (e) The upper figure shows the way to cut an armchair ribbon with the weak bond ($t_1$) broken. The lower figure shows the finite-size armchair ribbon. (f) The energy spectrum of a finite ribbon in Fig. S6(e) when $t_2 > t_1$. (g) The way of cutting a 1D zigzag ribbon with strong bond ($t_2$) broken. (h) The energy spectrum of the zigzag ribbon in Fig. S6(g). The red box marks the dangling-bond states on zigzag edges. The lower figure shows the real-space distribution of the dangling-bond states. (i) The corner states of a rhombus sample. The grey and pink band mark the zigzag and armchair ribbon, respectively.

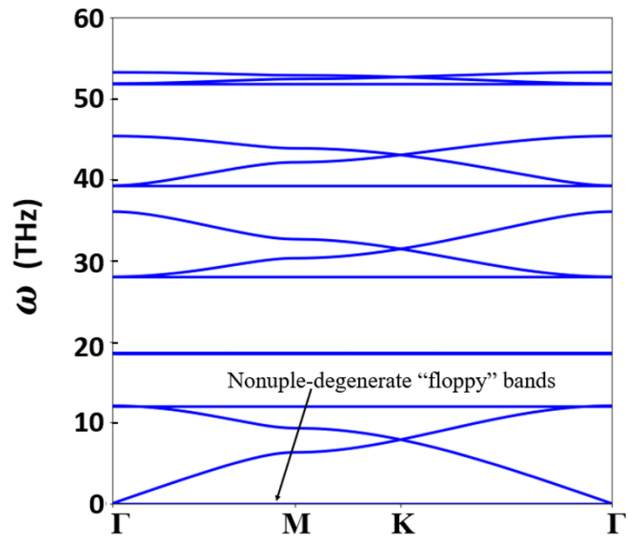

Fig. S7. In-plane phonon bands with four atoms on each Kagome lattice site to model Kagome-BN. We note that there are total 24 bands, but only 15 are visible, because there is a nonuple-degenerate flat band at zero frequency, as indicated. They are so-called floppy modes [16,17], which appear when the number of spring constraints is less than the number of vibrational degrees of freedom. Here, we considered only the in-plane stretching but not bending restoring force for springs, leading to the floppy bending modes with zero frequency which can be stabilized at finite frequency by adding a weak bending restoring force.